\documentclass[fleqn,twoside]{article}
\usepackage{espcrc2}

\readRCS $Id: espcrc2.tex,v 1.2 2004/02/24 11:22:11 spepping Exp $
\usepackage[dvips]{graphicx}
\usepackage[figuresright]{rotating}
\newcommand{\be}{\begin{equation}}
\newcommand{\ee}{\end{equation}}
\newcommand{\ba}{\begin{array}}
\newcommand{\bqa}{\begin{eqnarray}}
\newcommand{\eqa}{\end{eqnarray}}

\newcommand{\AmS}{{\protect\the\textfont2
  A\kern-.1667em\lower.5ex\hbox{M}\kern-.125emS}}
\title{Update analysis of $\tau^- \to V P^- \nu_\tau$: Theory and Experiment}
\author{Zhi-Hui Guo
  \address{ \mbox{ Department of Physics, Peking University, Beijing 100871, P.~R.~China, } }
  \address{\mbox{ IFIC, CSIC-Universitat de Val\`encia, Apt.~Correus 22085, E-46071, Val\`encia, Spain} }
  \thanks{
I would like to thank the organizers of QCD 2008 for providing the
charming atmosphere in the conference and also the nice social
events. ZHG is supported in part by China Scholarship Council and
National Nature Science Foundation of China under grant number
10721063 and 10575002. This work is also partially supported by EU
Contract No. MRTN-CT-2006-035482 (FLAVIAnet), by MEC (Spain) under
grant FPA2007-60323 and by Spanish Consolider-Ingenio 2010 Programme
CPAN (CSD2007-00042).}  }

\begin{document}

\begin{abstract}
Within the resonance chiral theory (R$\chi$T), we have studied the
process of a tau lepton decaying into a vector resonance plus a
pseudo-Goldstone meson and a tau neutrino. Two kinds of processes
are discussed: (a) $\tau^-\to(
\rho^0\pi^-,\omega\pi^-,\phi\pi^-,K^{*0}K^-)\nu_\tau$, belonging to
$\Delta S=0$ processes and (b) $\Delta S = 1$ processes, such as
$\tau^- \to (\rho^0 K^-,\omega K^-,\phi K^-,\overline{K}^{*0}\pi^-
)\nu_\tau$. To fit the $\tau^- \to \omega \pi^- \nu_\tau$ spectral
function and the decay distribution of $\tau^- \to \omega K^-
\nu_\tau$ to get unknown resonance couplings, we then make a
prediction for branching ratios of all channels.
\end{abstract}

\maketitle
\section{Introduction}
Due to its clean background, $\tau$ decay can provide an excellent
environment to study the nonperturbative dynamics of QCD. In these
decays, the intermediate resonances may play an important role.
Moreover, due to the improvement of statistically significant
measurements, the branching ratios and spectral functions of the
processes containing resonances in the final states of $\tau$ decays
have also been determined in recent experiments
\cite{pdg}\cite{expwpi}\cite{expwk}\cite{expphikbelle}\cite{expphikbabar}.
Motivated by the new measurements, we perform the study of tau
decaying into a resonance plus a pseudo-Goldstone meson and a tau
neutrino in this work.

Combining the $ \rm{SU(3)_L} \times \rm{SU(3)_R}$
chiral symmetry, that drives the interaction of
pseudo-Goldstone mesons resulting from the spontaneous chiral
symmetry breaking of QCD and the $\rm{SU(3)_V}$ symmetry for the resonance
multiplets, the resonance chiral lagrangians consisting of specific number
of resonance multiplets have been written down in
\cite{rchpt89}\cite{rchptext}. To build a more realistic QCD-like effective theory,
large-$N_C$ techniques and short-distance constraints from QCD have  been implemented into
the resonance effective theory to constraint resonance couplings
\cite{moussallam}\cite{knecht}\cite{vvp}\cite{vap}.
Therefore, resonance chiral effective theory can be a perfect tool to study
hadronic $\tau$ decays. Indeed it has already been employed in the
studies of $\tau \rightarrow \pi K \nu_\tau$ \cite{taupik}, $\tau
\rightarrow \pi\pi\pi\nu_\tau$ \cite{tau3pi} and $\tau \rightarrow
K\overline{K}\pi\nu_\tau$ \cite{tau2kpi}.

In Ref.\cite{tauvp}, we have made a comprehensive analysis for tau
decaying into a vector resonance plus a pseudo-Goldstone meson and a
tau neutrino: (a) $\Delta S =0$ processes, such as $\tau^-
\rightarrow ( \rho^0\pi^-, \omega\pi^-, \phi\pi^-, K^{*0}K^- )
\nu_\tau$ and (b) $\Delta S = 1$ processes, like $\tau^- \rightarrow
( \rho^0 K^-, \omega K^-, \phi K^-, \overline{K^{*0}}\pi^-)
\nu_\tau$, in the frame of R$\chi$T. The main results will be presented in
this paper.

\section{ Theoretical frame for tau decays }

The amplitude for $\tau^-(p) \rightarrow P^-(p_1) V(p_2)
\nu_\tau(q)$, where $P^-$ can be $\pi^-, K^-$ and $V$ can be
$\rho^0, \omega, \phi, K^{*0}, \bar{K^{*0}}$, has the general
structure
\bqa\label{general}
&& -G_F V_{uQ} \overline{u}_{\nu_\tau}(q)
\gamma^\mu(1-\gamma_5)u_{\tau}(p)\,\nonumber \\ &&
\times \epsilon^{*\nu}_V(p_2)\big[\, v\, \varepsilon_{\mu\nu\rho\sigma} p_1^{\rho}
p_2^{\sigma}\nonumber \\ &&-(a_1g_{\mu\nu} + a_2p_{1\mu} p_{1\nu}+a_3p_{2\mu}p_{1\nu})
\,\big],
\eqa
 where $G_F$ is the Fermi constant; $V_{uQ}$ is the CKM matrix
element; $\varepsilon_{\mu\nu\rho\sigma}$ is the anti-symmetric
Levi-Civit\`a tensor; $\epsilon^{*\mu}_V(p_2) $ is the polarization
vector for the vector resonance; $v$ denotes the form factor of the
vector current
 and $a_1, a_2, a_3$ are the corresponding axial-vector form factors.

The form factors in Eq.(\ref{general}) will be calculated within
R$\chi$T. The relevant R$\chi$T lagrangian that we will use in this
paper is
{\small \bqa
&&\mathcal{L}_{R\chi T}= \frac{F^2}{4}\langle u_\mu u^\mu + \chi_+ \rangle
+\mathcal{L}^{kin}(V, A)+\mathcal{L}_{2V, A}
\nonumber \\ &&
+\mathcal{L}_{VVP}+\mathcal{L}_{VJP}+\mathcal{L}_{VAP}
++\mathcal{L}_{V{V_1}P},
\eqa}
where the first term is the leading order operators of the chiral perturbation
theory \cite{gl845}; the antisymmetric tensor formalism will be used to describe
the vector and axial-vector resonances; the kinematics terms
$\mathcal{L}_{kin}(V, A)$ and the operators only containing one multiplet of
resonances $\mathcal{L}_{2V, A}$ can be found in \cite{rchpt89};
the operators containing two resonance multiplets:
$\mathcal{L}_{VVP}, \mathcal{L}_{VJP}, \mathcal{L}_{VAP}$,
can be found in \cite{vvp}\cite{vap}; the interaction terms between
the lowest vector multiplet $V$ and the heavier multiplet $V_1$ can be found
in \cite{rhoprime}. The SU(3) matrices for vector and axial-vector
resonances are given by
\bqa
{\tiny V= \left(
\begin{array}{ccc}
\frac{\rho_0}{\sqrt2}+\frac{\omega_8}{\sqrt6}+\frac{\omega_1}{\sqrt3} & \rho^+ & K^{*+} \nonumber \\
\rho^- & -\frac{\rho_0}{\sqrt2}+\frac{\omega_8}{\sqrt6}+\frac{\omega_1}{\sqrt3}  & K^{*0} \nonumber \\
K^{*-} & \overline{K}^{*0} &-\frac{2\omega_8}{\sqrt6}+ \frac{\omega_1}{\sqrt3} \nonumber \\
\end{array}
\right), }
\eqa
\bqa
{\tiny A= \left(
\begin{array}{ccc}
\frac{a_1^0}{\sqrt2}+\frac{f_{1}^8}{\sqrt6}+\frac{f_{1}^1}{\sqrt3} & a_1^+ & K_{1A}^{+} \nonumber \\
a_1^- & -\frac{a_1^0}{\sqrt2}+\frac{f_{1}^8}{\sqrt6}+\frac{f_{1}^1}{\sqrt3}  & K_{1A}^{0} \nonumber \\
 K_{1A}^{-} & \overline{K}_{1A}^{0} & -\frac{2f_{1}^8}{\sqrt6}+\frac{f_{1}^1}{\sqrt3} \nonumber \\
 \end{array}
 \right),}
\eqa
and $K_{1A}$ is related to the physical states $K_{1}(1270),
K_{1}(1400)$ through:
\bqa\label{theta} K_{1A} &=& \cos\theta\,\,
K_{1}(1400) + \sin\theta \,\,K_{1}(1270) \,.
\eqa
About the discussion on the nature of $K_1(1270)$ and $K_{1}(1400)$,
one can see  \cite{thetasuzuki} for details.
For the vector resonances $\omega$ and $\phi$, we
assume the ideal mixing for them throughout:
\bqa
\omega_1 = \sqrt{\frac{2}{3}} \omega - \sqrt{\frac{1}{3}} \phi, \quad
\omega_8 = \sqrt{\frac{2}{3}} \phi + \sqrt{\frac{1}{3}} \omega \,.
\eqa

\begin{figure}[ht]
\begin{center}
\includegraphics[angle=0, width=0.43\textwidth]{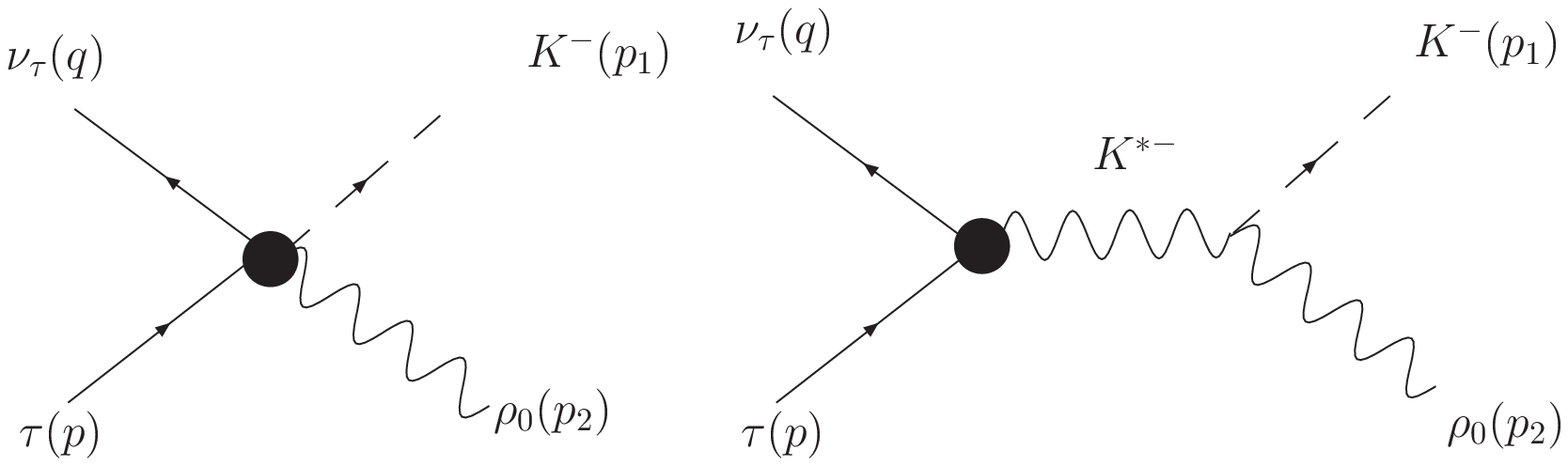}
\caption{Diagrams appearing in the vector current of $\tau^-(p) \to K^-(p_1) \rho^0(P_2) \nu_\tau(q)$.
\label{figv} }
\end{center}
\end{figure}

\begin{figure}[ht]
\begin{center}
\includegraphics[angle=0, width=0.43\textwidth]{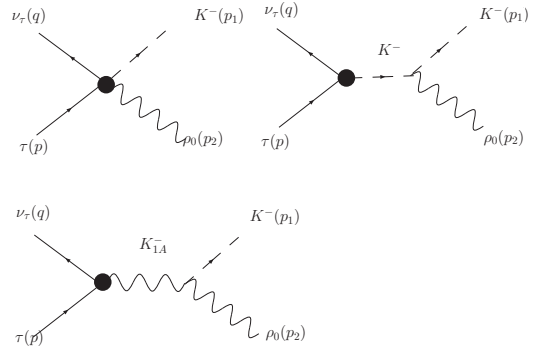}
\caption{Diagrams appearing in the axial-vector current of $\tau^-(p) \to K^-(p_1) \rho^0(P_2) \nu_\tau(q)$. \label{figa} }
\end{center}
\end{figure}

The corresponding Feynman diagrams contributed
to the form factors $v,a_1, a_2, a_3$ in the process of
$\tau^-(p) \rightarrow  K^-(p_1) \rho^0(p_2)\nu_\tau(q)$ are given in
Fig.(\ref{figv}) and Fig.(\ref{figa}).

The explicit expressions for the form factors $v,a_1, a_2, a_3$
of $\tau^-(p) \rightarrow  K^-(p_1) \rho^0(p_2)
\nu_\tau(q)$  can be found in \cite{tauvp}. The corresponding
form factors of other channels are quite similar to the ones of
$\tau^-(p) \rightarrow  K^-(p_1) \rho^0(p_2)\nu_\tau(q)$ and one can find
the explicit expressions in the Appendix of \cite{tauvp}.

Since most of the intermediate resonances have wide decay
widths, the off-shell widths of these resonances may play an
important role in the dynamics of $\tau$ decays. To introduce the
finite decay widths for the resonances implies that the corrections
from the next-to-leading order of $1/N_C$ expansion are taken
account into our game. This issue has been discussed in
\cite{engwidthrho} for the decay width of $\rho(770)$ and we take
the result of that article
{\small \bqa \label{gammarho} \Gamma_\rho(s) =
\frac{s M_V}{96 \pi F^2}\left[ \sigma_\pi^3 \theta(s-4m_\pi^2) +
\frac{1}{2}\sigma_K^3 \theta(s-4m_K^2) \right],\nonumber
\eqa}
where
$\sigma_P =\sqrt{1-4m_P^2/s}$ and $\theta(s)$ is the step function.
About the energy dependent widths for $\rho', K^*, {K^*}',
K_1(1270), K_1(1400), a_1(1260)$, we follow the way
introduced in \cite{taupik} to construct them and
the explicit formulae can be found in \cite{tauvp}.
\begin{table*}
\begin{tabular}{ccccc}
 \hline\hline
 &  Exp. & One multiplet & Fit 1 & Fit 2 \\
\hline $B(\tau^-\rightarrow \rho^0\pi^-\nu_\tau)$\,&\,--- \,&\,
$8.1 \times 10^{-2}$\,
&\,$9.4 \times 10^{-2}$ \,&\,$8.1 \times 10^{-2}$\\
 $B(\tau^-\rightarrow \omega\pi^-\nu_\tau)$ \,&\, $(1.95 \pm 0.08) \times 10^{-2}$ \,
 &\, $0.17  \times 10^{-2}$\,&\, $ 2.1
\times 10^{-2}$ \,&\, $2.1  \times 10^{-2}$\\
$B(\tau^-\rightarrow K^{*0}K^-\nu_\tau)$\,&\, $(2.1 \pm 0.4)
\times 10^{-3}$ \, &\, $1.4 \times 10^{-3} $\,&\, $1.5 \times
10^{-3}$ \,&
\,$1.5 \times 10^{-3}$ \\
\hline\hline
\end{tabular}
\caption{\small Branching ratios for $\Delta S =0$ processes. The second
column denotes experimental values, which are taken from \cite{pdg}.
The values from the third column to the fifth column denote our
predictions under different assumptions: only including the lowest
multiplet, the fitting results with $\lambda'=0.455,\,\lambda''=-0.0938,\,\lambda_0=0.0904$ (Fit 1)
and the fitting results with
$\lambda' =0.5, \, \lambda'' =0, \, \lambda_0 = 0.125$ (Fit 2)\,. It is interesting to notice that
a recently reported number: $B(\tau^-\rightarrow K^{*0}K^-\nu_\tau)=(1.56 \pm 0.02 \pm 0.09)\times 10^{-3}$
\cite{taukvk}, is highly consistent with our predictions.
\label{tab1}}
\end{table*}

\begin{table*}
\begin{tabular}{ccccc}
\hline\hline
  & Exp & One multiplet & Fit 1 & Fit 2 \\
\hline $B(\tau^-\rightarrow \rho^0K^-\nu_\tau)$\,&\, $(1.6 \pm 0.6)
\times 10^{-3} $  \,
&\,  $3.9  \times 10^{-4} $ \,& \, $4.7  \times 10^{-4} $ \,&\, $3.5  \times 10^{-4}$ \\
 $B(\tau^-\rightarrow \omega K^-\nu_\tau)$\,&\,$(4.1 \pm 0.9)
\times 10^{-4}$  \, &\, $3.5  \times 10^{-4}$ \,&
\,$ 4.0  \times 10^{-4}$ \,&\, $3.0  \times 10^{-4}$ \\
 $B(\tau^-\rightarrow \phi K^-\nu_\tau)$\,&\, {\footnotesize
$(4.05 \pm 0.36)  \times 10^{-5}$(Belle)}\,
&\, $1.7 \times 10^{-5}$ \, &\,$1.8 \times 10^{-5}$ \,&\, $1.6 \times 10^{-5}$ \\
\,&\, {\footnotesize $(3.39 \pm 0.34)\times 10^{-5}$({\sc BaBar})} \,&\,  \,&\, \,&\,\\
 $B(\tau^-\rightarrow \overline{K}^{*0}\pi^-\nu_\tau)$\,&\,
$(2.2 \pm 0.5) \times 10^{-3}$  \,&
\, $3.3  \times 10^{-3}$ \,&\,$5.1  \times 10^{-3}$ \,&\, $4.0 \times 10^{-3}$ \\
\hline\hline
\end{tabular}
\caption{\small Branching ratios for $\Delta S =1$ processes. The meaning
of numbers in different columns is the same to Table \ref{tab1}. The
experimental values for $\phi K^-$ are taken from
\cite{expphikbelle} and \cite{expphikbabar}. The remaining
experimental data is taken from \cite{pdg}. \label{tab2}}

\end{table*}

\begin{figure}[!htb]
\begin{center}
\includegraphics[angle=0, width=0.45\textwidth]{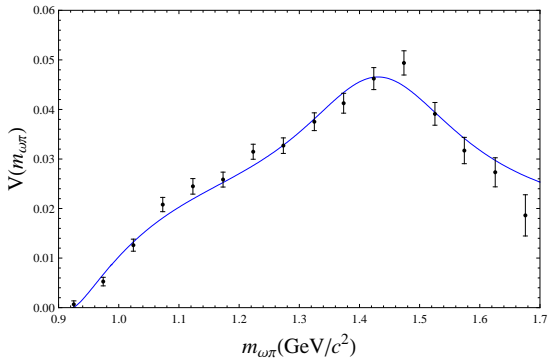}
\caption{Spectral function for
$\tau^-\rightarrow\omega\pi^-\nu_\tau$. The experimental data are
taken from \cite{expwpi}. \label{fig1} }
\end{center}
\end{figure}

\begin{figure}[!htb]
\begin{center}
\includegraphics[angle=0, width=0.45\textwidth]{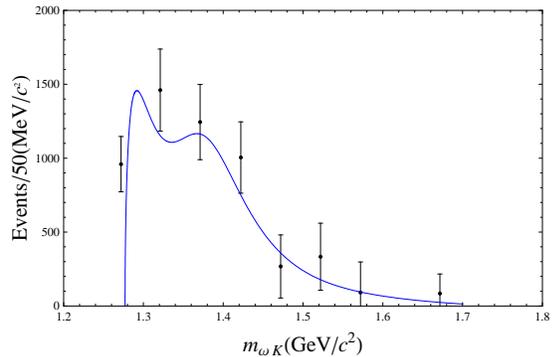}
\caption{Invariant mass distribution for $\omega K^-$ in the process
of $\tau^-\rightarrow\omega K^-\nu_\tau$. The experimental data are
taken from \cite{expwk}. \label{fig2}}
\end{center}
\end{figure}

\begin{figure}[!htb]
\begin{center}
\includegraphics[angle=0, width=0.45\textwidth]{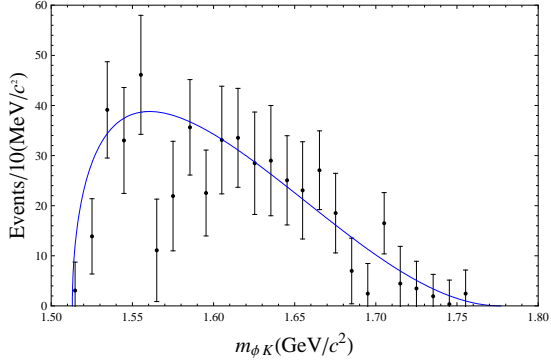}
\caption{Predicted invariant mass distribution for $\phi K^-$ in the process
of $\tau^-\rightarrow\phi K^-\nu_\tau$.
 The experimental data are taken from \cite{expphikbelle}, where only the data up to
 $m_{\phi K}=1.75$ GeV are quoted in the plot.\label{fig3}}
\end{center}
\end{figure}

\section{Phenomenological discussion}

Although some related resonance couplings can be fixed by
imposing QCD short distance constraints \cite{vvp}\cite{vap}\cite{tauvp},
we still have 5 free parameters: $d_3$, a resonance coupling
related to the lowest vector multiplet; $d_m, d_M, d_s$,
the couplings for the excited vector multiplet $V_1$;
the mixing angle $\theta$, a parameter for the axial-vector
resonances defined in Eq.(\ref{theta}). We fit the 5
parameters using the $\tau^-\rightarrow \omega\pi^- \nu_\tau$
spectral function and the invariant mass distribution of $\omega K$
system in the process of $\tau^-\rightarrow \omega K^-\nu_\tau$.
To test the stability, we take two sets of values for the axial-
vector resonance couplings $\lambda_i$ \cite{tauvp} in the fit.

Our fitting result for $\theta$, defined in Eq.(\ref{theta}),
is  $|\theta|=58.1^{+8.4}_{-7.3}$
degrees, which is consistent with
$|\theta| = 37^{\textrm{o}}$ and $58^{\textrm{o}}$
recently determined also in $\tau$ decays \cite{thetahycheng}.
The predictions for branching ratios we get are summarized in Table \ref{tab1} and Table
\ref{tab2}. In case of the one multiplet, we take the values of resonance couplings from
\cite{vvp}\cite{vap}. For the other cases, we use the fitting results presented
in \cite{tauvp}. Comparison of the figures we have obtained for the
$\tau^- \rightarrow \omega \pi^- \nu_\tau$ spectral function and the
invariant mass distributions for $\omega K^-, \phi K^-$ between the
experimental data are given in Fig.(\ref{fig1}-\ref{fig3})
respectively. Although different choices for $\lambda_i$ affect the
branching ratios, the invariant mass distributions are barely
influenced. So we only plot the figures with
 $\lambda'=0.5, \lambda''=0, \lambda_0=0.125$.

\end{document}